# DUSTY PLASMA CORRELATION FUNCTION EXPERIMENT

B. Smith, J. Vasut, T. Hyde, L. Matthews, J. Reay, M. Cook, J. Schmoke

*Center for Astrophysics, Space Physics and Engineering Research (CASPER),
Baylor University, P. O. Box 97310, Waco, TX 76798-7310, USA*

**ABSTRACT**

Dust particles immersed within a plasma environment, such as those in protostellar clouds, planetary rings or cometary environments, will acquire an electric charge. If the ratio of the inter-particle potential energy to the average kinetic energy is high enough the particles will form either a "liquid" structure with short-range ordering or a crystalline structure with long range ordering. Many experiments have been conducted over the past several years on such colloidal plasmas to discover the nature of the crystals formed, but more work is needed to fully understand these complex colloidal systems. Most previous experiments have employed monodisperse spheres to form Coulomb crystals. However, in nature (as well as in most plasma processing environments) the distribution of particle sizes is more randomized and disperse. This paper reports experiments which were carried out in a GEC rf reference cell modified for use as a dusty plasma system, using varying sizes of particles to determine the manner in which the correlation function depends upon the overall dust grain size distribution. (The correlation function determines the overall crystalline structure of the lattice.) Two dimensional plasma crystals were formed of assorted glass spheres with specific size distributions in an argon plasma. Using various optical techniques, the pair correlation function was determined and compared to those calculated numerically.

## INTRODUCTION

In recent years, the study of systems consisting of dust and plasmas has become especially significant. This is perhaps due to the almost universal existence of the combination of dust and plasma in systems ranging from protostellar clouds to plasma processing environments. Such complex plasmas also exist in planetary rings, cometary environments, and in the Earth's ionosphere and magnetosphere. Not only do dusty plasmas model naturally occurring systems like those above it is now thought that they can also mirror the structure, dynamics, and phase transitions of certain condensed matter physics. Therefore, an understanding of the processes involved in such systems is important.

Whether these systems are characterized as dusty plasmas, dust in plasma, colloidal plasmas, or crystalline plasmas, they are governed by the same basic physics. Dust immersed in a plasma will acquire an electric charge through the processes of ion and electron collisions, photoemission, and secondary electron emission. In low temperature plasmas, without ultraviolet light, secondary electron emission and photoemission can be ignored. Charging, therefore, depends only on the ion and electron currents. Usually, the charge obtained by a grain is negative, since charging is ruled by the incident flux of the faster electrons from the enveloping plasma (Morfill et al., 1999).

Colloidal plasmas can be loosely defined as a dusty plasma in which the dust can no longer be considered a contaminant but should instead be considered an integral component of the plasma. Structures called Coulomb crystals can form within colloidal plasmas, once certain criteria are met. The Coulomb coupling parameter ($\Gamma$), the ratio of interparticle potential energy to the particles' thermal energy, determines how ordered these structures can become. If $\Gamma$ is greater than 170 (Ichimaru, 1982), the system behaves like a solid, forming a crystalline lattice. In other words, long-range ordering is present. Thomas and colleagues experimentally observed such types of crystallization in 1994, since then Coulomb crystals have been observed in a variety of plasma environments. If short-range ordering is present in the system, the crystal is said to be in a liquid phase. Finally if the colloidal plasma exhibits no ordering the system is said to be in a gas phase.

Most experiments have used monodisperse spheres in single component plasmas to examine Coulomb crystals (e.g. Pieper et al., 1996a; Pieper et al., 1996b; and Zuzic et al., 2000). However, astrophysical environments, as well as many laboratory ones, have significant size variation within the dust distribution which probably affects the ordering present in the system. In this experiment, we seek to discover how a broader distribution of particle sizes affects the crystalline structure of a colloidal plasma. Recently, it has been theoretically shown that a system with a more diffuse particle distribution will form a more liquid like crystal (Vasut et al., 2002b).

**EXPERIMENT**

These experiments were carried out in a Gaseous Electronics Conference (GEC) radio frequency (rf) Reference Cell (Hargis et al., 1994). The upper electrode of the GEC rf Reference Cell was replaced with a grounded cylidrical electrode. To allow the dust crystal to be imaged from above, an optical window was installed in the upper electrode port with a feed-through for dust particles. The lower electrode was capacitively coupled to a rf generator and amplifier though a T-type matching network. The walls of the cell also serve as grounds. The lower electrode has a stainless steel sheath ground shield and a Teflon insulator. An aluminum dish with a circular cutout rests on the lower electrode. The cutout sets up boundary conditions which constrain the dust particles to remain above the electrode.

The discharge region was illuminated using a horizontal diode laser sheet and imaged using a CCD (charge-coupled device) camera mounted vertically above the cell. This camera contained a band pass filter to reduce ambient light created by the glow discharge plasma, allowing the viewing of the particles in the perpendicularly illuminated plane. To adequately resolve individual particles within the crystalline lattice, the top mounted camera used a zoom lens with a variable focal length of 18-108 mm, a six to one magnification ratio, and a lens magnification doubler, producing a field of view of approximately 11 mm by 11 mm. This camera system is ideal for observing the horizontal lattice structure.

A frame grabber board captures and transfers the data images to a computer. In this experiment the pictures were taken at a rate of 30 frames per second, interlaced. Each image had a resolution of 512 by 480 pixels and occupied about 250 kilobytes of disk space. These pictures were then analyzed via a MATLAB program (Boessé et al., 2002) in order to obtain the Voronoi diagrams and pair correlation functions needed to analyze the crystals.

The object of this study was to examine the level of ordering present when the particle distribution is not monodisperse. To that end, plasma crystals were formed from glass spheres having a 3 – 10 micron Gaussian distribution which was peaked at 7 microns. These particles were purchased from *Polysciences, Inc.*, and were quoted as having 60% of the particles between 6 and 8 microns. The experiments were performed at 1 Torr, the critical pressure for argon (Pieper et al. 1996a), with an applied voltage of 30 $V_{rms}$ in an argon plasma. At pressures of 1 Torr and above for the monodisperse particles used by Pieper et al. (1996a) "trapping was stable at all available powers." Collected data was then compared to data taken under the same conditions using 8.9-micron polymethylene melamine spheres.

The experimental results were compared to theoretical results from a computer model previously used to examine Coulomb crystallization (Vasut and Hyde, 2001). This computer model employs a Barnes-Hut tree code (1986) known as "Box_Tree" developed by Derek Richardson (1994). The code was later modified to include electrostatic forces and Debye shielding in order to better simulate dusty plasma environments (Matthews and Hyde, 1998; Matthews and Hyde, 2002; Swint and Hyde, 2002; Vasut et al., 2002a; Qiao and Hyde, 2002a; and Qiao and Hyde, 2002b). Box_Tree calculates all near-field forces directly, even as the code approximates medium- and far-field forces using multipole terms for collections of particles.

**RESULTS**

Visual inspection of the resulting crystals show that the glass spheres do not form as ordered a structure at 1 Torr compared to the crystal formed from 8.9 micron polymethylene melamine spheres. By stepping the horizontal laser vertically through the particle cloud, it was noted that the particle cloud had two vertically diffuse layers. For the purposes of this experiment, the bottom layer was chosen for analysis. While sedimentation of the larger particles was minor, it is believed that the bottom layer had a higher concentration of the largest particles. Samples of the raw images of the ordered system for both the glass and polymethylene melamine spheres are shown in Figure 1 and Figure 2 respectively. Although it is possible to approximate ordering directly from the images taken of the crystal, more accurate methods for determining the order present must be used before system classification can occur.

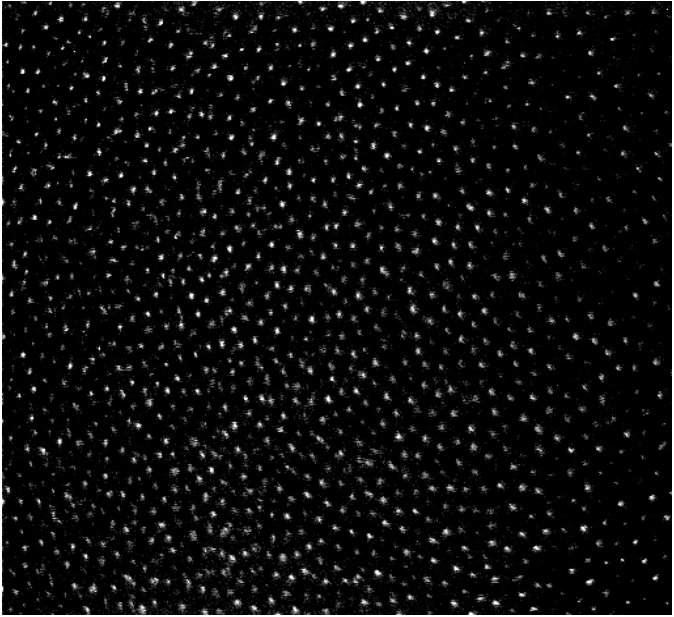

Fig. 1. Raw image of 3-10 micron glass spheres. The bright spots are the dust particles.

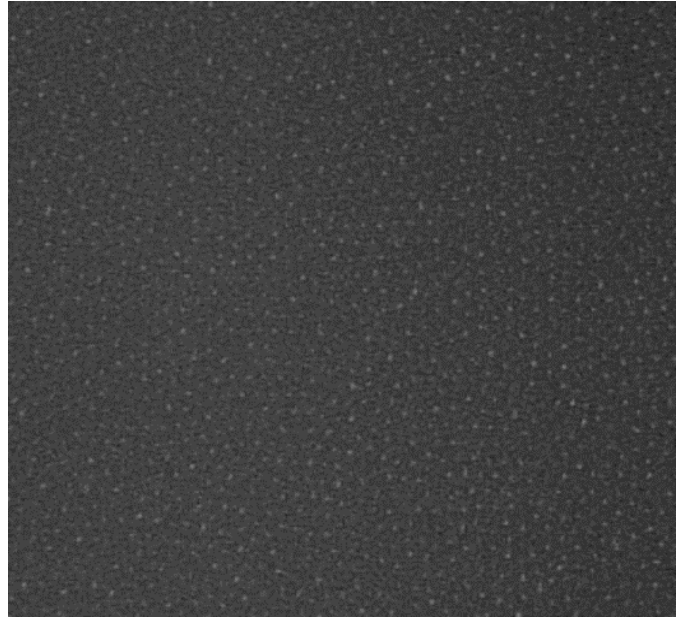

Fig. 2. Raw image of 9 micron polymethylene melamine spheres. The bright spots are the dust particles.

There are several methods for measuring the amount of order in a complex plasma system. One simple method is to find the number of nearest neighbors for a typical dust particle. In two dimensions this can be done by dividing any single plane into a series of cells, with the boundary between the cells comprised of points equidistant from two or more particles. Such a figure is known as a Voronoi diagram, an example of which is shown in Figure 3. In a perfect hexagonal lattice, each cell will be a hexagon with the six sides representing its six nearest neighbors. After the Voronoi diagram is generated, the average number of neighbors a cell has and/or the variation in that number can serve as an indication of overall order for the system. Similar diagrams and nearest neighbor calculations may be performed for three or higher dimensions, but in this work the standard Voronoi technique is used in which only two-dimensional cases are studied. As can be seen in Figure 3, the glass spheres do not form a solid crystalline lattice. However, over 80% of the cells are 5, 6, or 7-sided with over 40% of the cells being 6-sided; therefore, some degree of ordering is present.

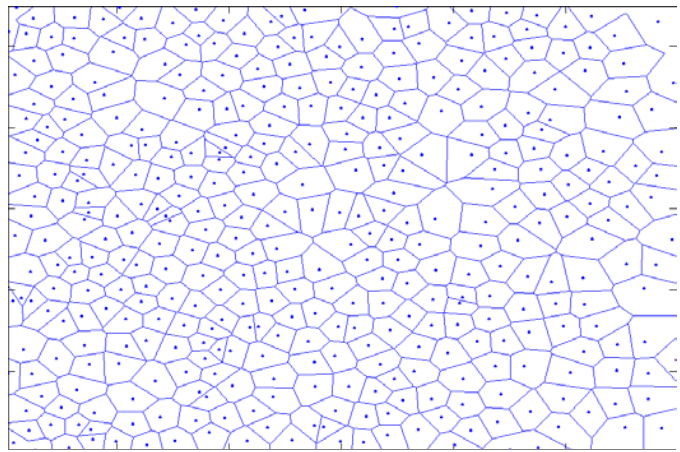

Fig. 3. Voronoi diagram for the image of 3-10 micron glass spheres shown in Figure 1. In this image, there are 94 5-sided cells, 157 6-sided cells, and 71 7-sided cells out of a total of 386 cells.

In order to more fully examine the crystalline structure of the Coulomb crystal, the pair correlation function, $g(r)$, is used. This function is also known as the radial or pair density distribution. This technique has been used in two-dimensional colloidal science analysis for some time and was first used to describe order within dust crystals by Quinn and colleagues (1996). The correlation function is most often used in two-dimensional studies, but has also been used in three dimensions (Zuzic, 2000) where it can help identify the type of lattice (body-centered cubic, face-centered cubic, or hexagonal closed-packed).

The pair correlation function represents the probability of finding two particles separated by a distance $r$. It is generated by considering each particle, measuring the distance to every other particle and then counting the number of particles a distance $r$ to $r + \delta r$ from the particle. This is repeated for every particle until an average value is determined which is then normalized by dividing by the annular area between $r$ and $r + \delta r$. This is sometimes normalized again by setting the asymptotic value equal to one, in which case it may be referred to as the normalized pair correlation function (Quinn et al., 1996). The correlation functions generated for this paper follow this practice.

For a perfect crystal with zero thermal energy, $g(r)$ will be a series of Delta functions. For a physical lattice at finite temperature, the Delta functions will spread out and overlap, as seen in Figure 4. The correlation function shown in Figure 4 exhibits several distinct peaks. The fact that so many peaks can be seen indicates system ordering to several times the nearest-neighbor distance. This indicates a long-range correlation among the particles and an overall crystalline structure. In Figure 4, the initial peak is much larger than the others. This is due to the fact that even in systems with long-range ordering, the correlation between nearest neighbors is stronger than for more distant pairs. Because of this stronger correlation, the first peak is significantly narrower causing a corresponding increase in the height of the peak.

In a liquid state, characterized by only short-range ordering, there is only one primary peak or a primary peak followed by a small second or third peak. In a gaseous state there is no correlation between the particles and the correlation function shows no clear peaks, although often the mutual repulsion of the particles prevents them from close approaches to each other. This can cause the correlation function to be zero for small distances instead of one, as in an ideal gas. Examples of the correlation functions for a gaseous, liquid, and crystalline state are shown in Figures 4-6.

The normalized correlation function for the Gaussian distribution of glass spheres is shown in Figure 7. The prominent first peak and absence of second or third peaks indicates a liquid crystal structure. Comparison with the pair correlation function obtained numerically (Figure 8) shows good agreement between theoretical calculations and experimental results.

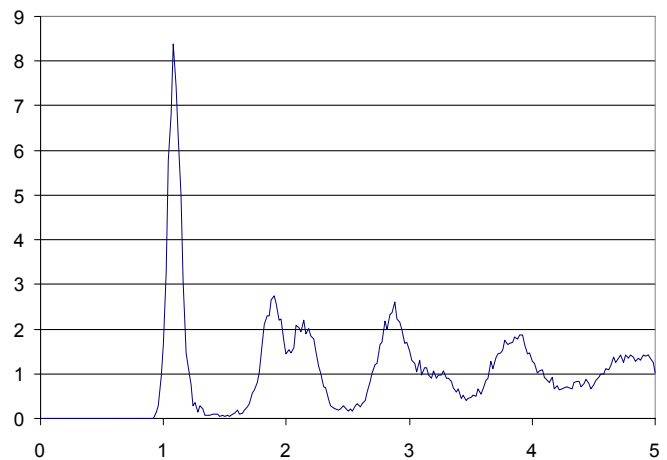

Fig. 4. Example of a normalized pair correlation function for a solid crystal.

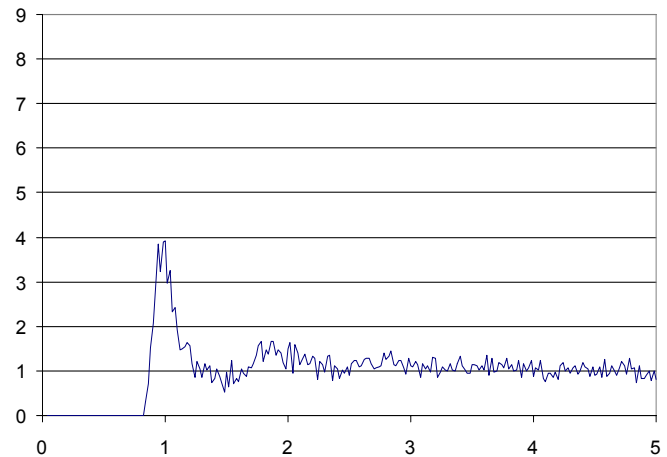

Fig. 5. Example of a normalized pair correlation function for a liquid crystal.

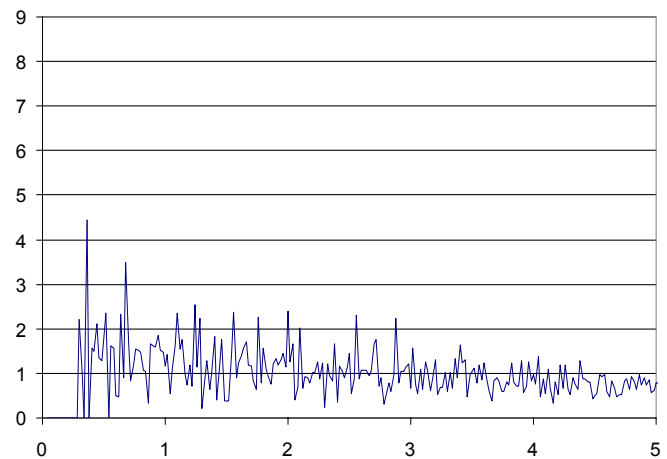

Fig. 6. Example of a normalized correlation function for a gaseous crystal.

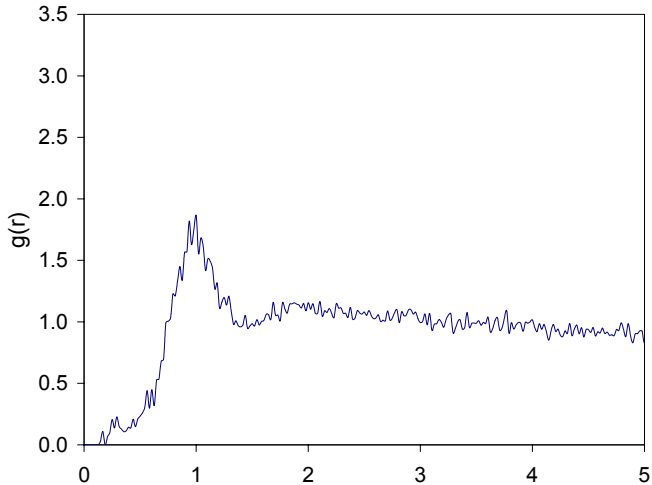

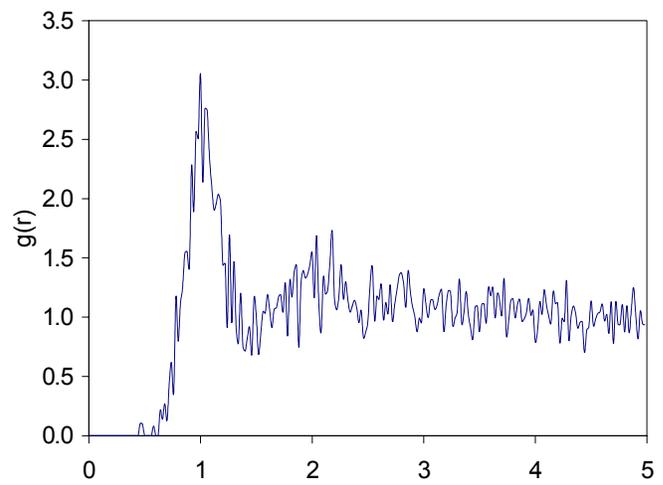

Fig. 7. Normalized pair correlation function for 3-10 micron glass spheres averaged over 10 frames. The x-axis has been normalized such that the first maximum occurs at one regardless of differences between successive frames.

Fig. 8. Normalized pair correlation function for numerical simulation. Particles were given a thirty-percent distribution in charge and mass.

## CONCLUSION

In this preliminary work, it has been shown that, while monodisperse particles will form solid Coulomb crystals, a Gaussian distribution of particles forms liquid Coulomb crystals instead. This is not a surprising result since Coulomb crystallization is particle charge dependent which in turn depends on the particle radius. However, the fact that a diffuse distribution forms a crystal at all is interesting since most naturally occurring systems will have some type of distribution in particle size. More work is necessary to fully explore the effects of a size distribution on Coulomb crystallization. This should include examining system ordering over additional power and pressure ranges and for various dust composition. Additionally, in situ methods of determining particle sizes within each layer should be included to examine the effects of the sedimentation of the larger particles.

## ACKNOWLEDGEMENTS

The writers gratefully acknowledge all those who have made the existence of CASPER possible, but would especially like to thank Vicky Santiago for her tireless work on this project. This research was funded in part by the National Science Foundation and the Office of Research at Baylor University.

## REFERENCES

Barnes, J., and P. Hut, A hierarchical O(N log N) force calculation algorithm, *Nature*, **324**, 446-449, 1986.

Boessé, C. M., M. K. Henry, T. W. Hyde, and L. S. Matthews, Digital imaging and analysis of dusty plasmas, submitted to *Adv. Space Res.*, 2002.

Morfill, G. E., H. M. Thomas, U. Konopka et al., The plasma condensation: liquid and crystalline plasmas, *Phys. Plasmas*, **6**(5), 1769-1780, 1999.

Hargis, P. J., K. E. Greenberg, P. A. Miller et al., The Gaseous Electronic Conference radio-frequency reference cell: A defined parallel-plate radio-frequency system for experimental and theoretical studies of plasma-producing discharges, *Rev. Sci. Instrum.*, **65**(1), 140-154, 1994.

Ichimaru, S., Strongly coupled plasmas: High density classical plasmas and degenerate electron liquids, *Rev. Mod. Phys.*, **54**, 1017-1059, 1982.

Matthews, L. S., and T. W. Hyde, Gravitoeletrodynamics in Saturn's F ring: Encounters with Prometheus and Pandora, submitted to *Journal of Physics A*, 2002.

Matthews, L. S., and T. W. Hyde, Numerical simulations of gravitoelectrodynamics in dusty plasmas, in *Strongly Coupled Coulomb Systems*, edited by T. Kalman et al., Plenum Press, New York, USA, 199-202, 1998.

Pieper, J. B., J. Goree, and R. A. Quinn, Experimental studies of two-dimensional and three-dimensional structure in a crystallized dusty plasma, *J. Vac. Sci. Technol. A*, **14**, 519-524, 1996a.

Pieper, J. B., J. Goree, and R. A. Quinn, Three-dimensional structure in a crystallized dusty plasma, *Physycal Review E*, **54**(5), 5636-5640, 1996b.

Qiao, K., and T. W. Hyde, Numerical simulation and analysis of thermally excited waves in plasma crystals, submitted to *Adv. Space Res.*, 2002.

Qiao, K., and T. W. Hyde, Dispersion relations for thermally excited waves in plasma crystals, submitted to *Adv. Space Res.*, 2002.

Quinn, R. A., C. Cui, J. Goree et al., Structural Analysis of a Coulomb Lattice in a Dusty Plasma, *Physical Review E*, **53**, R2049-R2052, 1996.

Richardson, D., Tree code simulations of planetary rings, *Mon. Not. R. Astron. Soc.*, **269**, 493-511, 1994.

Swint, G. S., and T. W. Hyde, Modeling chondrule melting using a resizing Box_Tree code, *Adv. Space Res.*, **29**(9), 1311-1316, 2002.

Thomas, H., G. E. Morfill, V. Demmel et al., Plasma crystal: Coulomb crystallization in a dusty plasma, *Physical Review Letters*, **73**(5), 652-655, 1994.

Vasut, J. A., and T. Hyde, Computer simulations of Coulomb crystallization in a dusty plasma, *IEEE Transactions on Plasma Science*, **29**(2), 231-237, 2001.

Vasut, J. A., M. D. Lennek, and T. W. Hyde, Plasma condensation and the one component plasma model, *Adv. Space Res.*, **29**(9), 1295-1300, 2002a.

Vasut, J. A., T. W. Hyde, and L. Barge, Finite Coulomb crystal formation, submitted to *Adv. Space Res.*, 2002b.

Zuzic, M., A. V. Ivlev, J. Goree et al., Three-dimensional strongly coupled plasma crystal under gravity conditions, *Physical Review Letters*, **85**(19), 4064-4067, 2000.